\documentclass[11pt,fleqn]{article}

\usepackage{amsfonts}
\usepackage{amsmath}
\usepackage{amssymb}
\usepackage{mathtools}
\usepackage{mathrsfs}
\usepackage{enumerate}
\usepackage{bbm}
\usepackage{graphicx,epsfig,amsthm,color}
\usepackage{pstricks}
\usepackage{subfig}
\usepackage{graphicx}
\usepackage{units}

\usepackage{caption}
\usepackage{booktabs}
\usepackage{textcomp}
\usepackage{array}
\usepackage{cite}
\usepackage{cancel}

\date{\today}

\newcolumntype{z}[1]{>{\RaggedRight\hspace{0pt}}p{#1}}
\newcolumntype{w}[1]{>{\RaggedRight\hspace{0pt}}p{#1}}
\newcolumntype{v}[1]{>{\Centering\hspace{0pt}}p{#1}}

\def\be{\begin{equation}}
\def\ee{\end{equation}}
\def\bea{\begin{eqnarray}}
\def\eea{\end{eqnarray}}

\def\be{\begin{equation}}
\def\ee{\end{equation}}
\def\bea{\begin{eqnarray}}
\def\eea{\end{eqnarray}}

\def\erp2{{\rm e}^{2\rho}}
\def\erm2{{\rm e}^{-2\rho}}
\def\er4{{\rm e}^{4\rho}}

\def\be{\begin{equation}}
\def\ee{\end{equation}}
\def\bea{\begin{eqnarray}}
\def\eea{\end{eqnarray}}

\def\m0{m_{\nu_{0,i}}}

\def\T0{T_{\nu_0}}

\newcommand{\half}{\frac{1}{2}}

\newcommand{\beqa}{\begin{eqnarray}}
\newcommand{\eeqa}{\end{eqnarray}}
\newcommand{\bpr}{\begin{problem}}
\newcommand{\epr}{\end{problem}}
\newcommand{\bcent}{\begin{center}}
\newcommand{\ecent}{\end{center}}
\newcommand{\bfig}{\begin{figure}}
\newcommand{\efig}{\end{figure}}
\newcommand{\bpc}{\begin{picture}}
\newcommand{\epc}{\end{picture}}

\renewcommand{\and}{A_{0}^{\nu ,D}(s)}

\newcommand{\bee}{\begin{equation}}

\def\beq{\begin{eqnarray}}
\def\eeq{\end{eqnarray}}

\newcommand{\bright}{\begin{flushright}}
\newcommand{\eright}{\end{flushright}}
\newcommand{\bminip}{\begin{minipage}}
\newcommand{\eminip}{\end{minipage}}

\DeclareCaptionLabelSeparator{mysep}{\hspace{3pt}:\hspace{3pt}}
\DeclareCaptionLabelFormat{mypiccap}{Fig.\hspace{3pt}{#2}}
\DeclareCaptionLabelFormat{mytabcap}{Tab.\hspace{3pt}{#2}}

\begin{document}

\date{}
\title{
\vskip 2cm {\bf\huge Quantum mechanics before the big bang in heterotic-M-theory}\\[0.8cm]}

\author{{\sc\normalsize Andrea Zanzi\footnote{E-mail: andrea.zanzi@unife.it} \!
\!}\\[1cm]
{\normalsize Dipartimento di Fisica e Scienze della Terra, Universit\'a di Ferrara - Italy}\\
[1cm]
}
\maketitle \thispagestyle{empty}
\begin{abstract}
{In this letter we investigate the role played by quantum mechanics before the big-bang in heterotic-M-theory assuming an orbifold compactification of time. As we will see particles are localized around a black hole but only in regions where a constructive quantum interference takes place. We infer that the creation of this interference pattern is interesting for many reasons: (A) it is a mechanism to localize particles on $S^4$ branes; (B) the Casimir potential for the dilaton can be interpreted as a gravitational effective potential for a two-body problem; (C) the quantum interference is a new way to define the branes in heterotic-M-theory. Remarkably, a modified Schroedinger equation is obtained. The stabilization of the branes' position is related to the absence of a cosmological singularity.}
\end{abstract}
%PACS numbers: 04.60.Cf, 98.80.-k, 95.36.+x

\clearpage

%\tableofcontents
\newpage

%%%%%%%%%%%%%%%%%%%%%%%%%%%%%%%%%%%%%%%%%%%%%%%%%%%%%%%%%%%%%%%%%%%%%%%%%%%%%%%%%%%%%%%%%%%%%%%%%%%%%%%%%%%%%%%%%%%%%
% Section INTRODUCTION
%%%%%%%%%%%%%%%%%%%%%%%%%%%%%%%%%%%%%%%%%%%%%%%%%%%%%%%%%%%%%%%%%%%%%%%%%%%%%%%%%%%%%%%%%%%%%%%%%%%%%%%%%%%%%%%%%%%%%
\setcounter{equation}{0}
\section{Introduction}

It is common knowledge that our equations of standard $\Lambda$CDM cosmology give us a cosmological singularity (a ''big-bang''). Many physicists interpret this singularity as the signal that we are exploiting the theory in a region of lengths and energies where the theory is not dependable anymore. Let us consider for example a homogeneous shell of spherical dust in the absence of angular momentum and let us suppose that gravity is the only interaction that we switch on in the theory. Needless to say, the spherical shell collapses under the effect of gravity and the matter density is an increasing function of time. In the laboratory, at short distances, interactions different from gravity will start to play a crucial role and a minimum radius will be reached after some time. If we insist in keeping only gravity in our theory, this minimum radius is not present anymore and an infinite matter density is predicted by the theory. This simple example shows that in order to avoid the singularity, we must introduce in the UV new interactions (or new degrees of freedom) which play a crucial role in making the matter density ''smooth''.

One of the more active research fields in modern theoretical physics is string cosmology\footnote{For an introduction to string theory the reader is referred to \cite{Tong:2009np, Blumenhagen:2013fgp} and references therein. For a textbook of standard cosmology the reader is referred to \cite{Weinberg:2008zzc}. Recent books of string cosmology are \cite{Erdmenger:2009zz, Gasperini:2007zz, Baumann:2014nda}. A very good resource of string cosmology is the M. Gasperini's website http://www.ba.infn.it/~gasperin/.}. The possibility of a pre-big-bang (PBB) phase has been analyzed in detail exploiting symmetries and dualities of string theory (see \cite{Gasperini:1992em, Gasperini:2002bn}). In the PBB scenario it is possible to obtain examples of bouncing cosmology. ''Bounce'' means a transition from a contracting to an expanding phase.

In this paper, in harmony with the analysis of \cite{Gasperini:1992em}, we imagine that a PBB phase is present. Formally, the pre (post) big bang phase can be identified with the string (Einstein) frame of our model, but, as already mentioned in \cite{Zanzi:2016het}, we are basically working in a unique conformal frame: the frame where masses are field dependent. From the theoretical point of view, our scenario will be the 5-dimensional heterotic-M-theory of references \cite{Lukas:1998tt, Brax:2002nt, Zanzi:2016het} where a naked singularity is predicted in the bulk from the supergravity (SUGRA) equations. The 5th dimension is compactified on a $S^1/Z_2$ orbifold. This naked singularity is a curvature singularity (i.e. a black hole) and the gravitational field is very strong near the black hole. We are particularly interested in a spherical collapse of matter (and radiation) into the black hole just before the big bang, namely when the universe was very small. The collapse we are considering takes place in the 5th dimension (radially) and all the matter of the universe takes part in the collapse towards the black hole (this radial collapse might be related, for example, to a contracting phase of the universe before the big bang).  Near the big bang, the universe is small, the scale  of the fifth dimension is planckian and it is natural to imagine that just before the big bang, there was a time when the size of the universe was comparable to the de Broglie wavelength of matter particles. In this peculiar configuration, quantum mechanics becomes important and quantum phenomena like the collapse of the wave function or the quantum interference cannot be neglected.

Recently in \cite{Zanzi:2016het}, we showed that, in the framework of heterotic-M-theory, it is possible to obtain the so called Modified Fujii's Model - MFM. In the MFM, namely the model that we are going to consider in this article, the collapse of the wave function of quantum mechanics has been analyzed in reference \cite{Zanzi:2015evj}. In this scenario the collapse of the wave function is a quantum gravity effect and it is induced by the chameleonic nature of the model (as far as chameleon fields are concerned, the reader is referred to \cite{Khoury:2013yya} for a review and to \cite{Khoury:2003aq, Khoury:2003rn} for the original proposal). One of the experiments that was discussed in detail in \cite{Zanzi:2015evj} was the diffraction of electrons through a circular hole. It is common knowledge that a typical pattern of interference with axial symmetry will be generated on the screen: some circular regions will be populated by electrons and other regions (where the interference is destructive) will not.

One of the purposes of this article is to show that $S^4$ branes can be generated around the black hole exploiting a constructive quantum interference near the big bang. This point must be further elaborated. In the chameleonic wave function collapse of \cite{Zanzi:2015evj}, there is no difference between a hole in a screen and a spherical shadow in a photon background. When we analyze the radial collapse of matter into the black hole, at a certain point matter particles will enter into the shadow of the black hole, this shadow is spherical and the chameleon field will perceive this shadow like a circular hole in a screen. The gravitational field of the black hole plays the role of a second screen (where the results of our quantum ''measurement of position'' are collected) because, once again, in the chameleonic collapse of the wave function there is no difference between a true screen and a region where the gravitational field is particularly intense. The length scales that we are considering are comparable to the de Broglie wavelengths because the universe is very small near the big bang. We infer that the collapse of matter into the spherical black hole just before the big bang is completely analogous to a diffraction of matter through a circular hole. The result is well known: there will be spherical regions where the quantum interference is constructive. These regions will be $S^4$ branes near the black hole.

The following scenario is taking shape. The quantum interference generates the $S^4$ branes. As usual, the positions of these branes are parametrized by moduli fields in the effective action. In principle these moduli can have a non-trivial dynamical behaviour, because the region of constructive interference does not necessarily correspond to a minimum of the moduli potential. Hence, quantum mechanics provides the initial position of the branes but the moduli potential is related to the dynamical evolution of the branes. This cosmological evolution of the branes' moduli (together with the stability of the branes) will be analyzed in a future work. At this stage, we simply remember that a stabilizing potential for these moduli can be obtained exploiting Casimir energy of bulk fields (see \cite{Zanzi:2006xr, Zanzi:2016het}). The final scenario is given by a set of $S^4$ branes rotating around the black hole (in harmony with \cite{Zanzi:2016het}). The radius of the branes is stable because the Casimir energy, related to $\alpha'$-corrections of the string \cite{Zanzi:2016het}, provides a repulsive contribution at short distance. 

Now one interesting point can be mentioned. We avoid the cosmological singularity because the Casimir stabilization guarantees a fixed distance (measured in the 5th dimension) between the branes and the black hole. Hence, the universe has a non-vanishing minimum size and all the fields are stabilized. This point must be further discussed. The Casimir-induced stabilization of the dilaton $\phi$ guarantees also the stabilization of matter fields $\Phi$, because the presence of the term $\phi^2 \Phi^2$ in the lagrangian of the MFM links $\phi$ to $\Phi$ in the minimum of the effective potential. In this way, when $\phi$ is stabilized, all the fields are stabilized and have a constant value. As already pointed out in \cite{Zanzi:2012du}, when all the fields are stabilized, the 4D curvature is constant and, consequently, the cosmological singularity is avoided. Summarizing, the gravitational collapse does not produce a cosmological singularity because, in our model, at short distances quantum mechanics and stringy effects are important.

One more remark is in order. In reference \cite{Zanzi:2015evj}, the collapse of the wave function has been analyzed but a modified Schroedinger equation was missing. In this article we fill this gap exploiting a peculiar compactification of time. Indeed, as already mentioned in \cite{Zanzi:2016het}, in the deep UV and deep IR regions, we can exchange space with time and, consequently, we can obtain a $S^1/Z_2$ orbifold of time. One of the consequences of this orbifold of time is that we can use the string frame equations even in the post big bang phase. This will be the key to write the modified Schroedinger equation in the MFM.
We point out that this orbifold of time is important also for the formation of $S^4$ branes: the analysis of \cite{Zanzi:2015evj} is based on a chameleonic behaviour of the dilaton after the big bang. Are we sure that the collapse of the wave function of reference \cite{Zanzi:2015evj} can be exploited before the big bang? The answer is that the orbifold of time identifies the pre with the post big bang phase. Therefore the chameleonic wave function collapse of \cite{Zanzi:2015evj} can be exploited before the big bang.

As far as the organization of this paper is concerned, in section 2 we summarize some useful ideas already discussed in the literature. The original results of this paper are presented in sections 3 and 4. In section 3 we discuss our mechanism to generate the $S^4$ branes. Section 4 presents the modified Schroedinger equation of the MFM. In the final section we draw some concluding remarks.

\section{Useful ideas from the literature}

In this section we will gather some results from the literature that will be very useful in the remaining part of this article. This section does not contain original results.

\subsection{The M-theory model}

The lagrangian we are interested in is the one of the so called Modified Fujii's Model - MFM. It contains two sectors, a scale invariant one and a symmetry breaking one.
Here is the scale invariant one:
\begin{equation}
{\cal L}_{\rm SI}=\sqrt{-g}\left( \half \xi\phi^2 R -
    \half\epsilon g^{\mu\nu}\partial_{\mu}\phi\partial_{\nu}\phi -\half g^{\mu\nu}\partial_\mu\Phi \partial_\nu\Phi
    - \frac{1}{4} f \phi^2\Phi^2 - \frac{\lambda_{\Phi}}{4!} \Phi^4 - \frac{\lambda_{\phi}}{4!} \phi^4
    \right).
\label{bsl1-96}
\end{equation}
$\phi$ is the string frame dilaton. $\Phi$ is another scalar field representative of matter fields. $f$, $\epsilon$, $\xi$, $\lambda_{\Phi}$ and $\lambda_{\phi}$ are constants. The symmetry breaking sector is due to the Casimir energy of bulk fields (see also \cite{Zanzi:2006xr}) and is made of a stabilizing potential for the dilaton. This symmetry breaking lagrangian breaks scale invariance explicitly and abundantly because we do not fine tune the parameters of the theory.

This MFM lagrangian has been discussed, in connection to (A) the cosmological constant problem in \cite{Zanzi:2010rs, Zanzi:2016het}, (B) heterotic-M-theory in \cite{Zanzi:2016het}, (C) the collapse of the wave function of quantum mechanics in \cite{Zanzi:2015evj}, (D) solar physics in \cite{Zanzi:2014aia}.

Let us summarize the M-theory origin of the MFM lagrangian.
Heterotic-M-theory is formulated in 11 dimensions. At intermediate energies, the theory can be described by a 5D braneworld model where the fifth dimension is compactified on a $S^1/Z_2$ orbifold. 3-branes are located at the fixed points of the orbifold and we can imagine that our universe is a portion of one of these 3-branes. In reference \cite{Zanzi:2016het} this scenario has been discussed with many 3-branes justified in the theory as M5 branes compactified on 2-cycles. 
For further details on heterotic-M-theory the reader is referred to \cite{Lukas:1998tt, Brax:2002nt}.
In 5D a dilaton field is present in the bulk. If no fields other than the bulk scalar field are present and no supersymmetry breaking is considered, then we are in a Bogomol'ny-Prasad-Sommerfield (BPS) configuration where the bulk equations coincide with the boundary conditions (there is a no-force condition for the branes in this configuration).
With the ansatz 
\bea
ds^2=dz^2+a(z)^2 g_{\mu\nu}dx^\mu dx^\nu
\eea

for the line element in 5D and with the exponential superpotential of the form ($k$ is constant)
\bea
U_B=4ke^{\alpha C}, \alpha \in \mathbb{R}
\eea
suggested by SUGRA, the BPS equations admits a dilatonic solution of the form
\bea
C(z)=-\frac{1}{\alpha}ln(1-4k \alpha^2 z)
\eea
while for the scale factor the solution is
\bea
a(z)=(1-4k \alpha^2 z)^{1/4\alpha^2}.
\eea
In the limit of small $\alpha$ we obtain an exponential scale factor 
\bea
a(z)=e^{-kz}.
\label{exp}
\eea
As we see from the divergent behaviour of the dilaton (which is related to a vanishing volume $V$ of the 6 extradimensions different from the orbifolded one) there is a naked singularity in the bulk (which is expected to be resolved when the full string theory description is taken into account). The singularity is screened by one of the two branes (the ''hidden brane''). The remaining brane can host the standard model particles and hence will be called the ''visible brane''.

A more realistic model can be obtained by putting matter on the branes and detuning the branes' tension. In this case at low energies the theory is described by a bi-scalar-tensor theory of gravitation, namely, two moduli participate into the description of gravity. These moduli are the dilaton $Q$ and the radion $R$. They are related to the position of a hidden brane (located near the singularity) and of a stack of 3-branes. To be more precise, two field redefinitions must be considered. The first redefinition is
\begin{eqnarray}
\tilde \phi^2 &=& \left(1 - 4k\alpha^2 z_{visible} \right)^{2\beta}, \label{posia1}\\
\tilde \lambda^2 &=& \left(1-4k\alpha^2 z_{hidden}\right)^{2\beta} \label{posia2},
\end{eqnarray}
with
\begin{equation}
\beta = \frac{2\alpha^2 + 1}{4\alpha^2}.
\end{equation}
The second redefinition is
\begin{eqnarray}
\tilde \phi &=& Q \cosh R, \label{posib1} \\
\tilde \lambda &=& Q \sinh R \label{posib2}.
\end{eqnarray}

The dilaton and the radion are both chameleon fields in this model\footnote{The chameleonic behaviour of the radion has been discussed in \cite{Brax:2004ym}.} and, consequently once the energy density of the environment is determined, the theory is basically described by a scalar tensor theory with a single modulus: the dilaton $\phi$ of the MFM lagrangian.

\subsection{The collapse of the wave function}

The MFM has been discussed in connection to the collapse of the wave function in quantum mechanics. In this paragraph we summarize some useful results from reference \cite{Zanzi:2015evj}.

When we perform a quantum measurement, we must interact with the quantum system. This interaction is related to an enhancement of the energy density of the environment. For example, in a Stern-Gerlach apparatus we exploit a magnetic field, in a diffraction experiment with electrons we exploit a screen characterized by a matter density larger than the region outside the screen. These considerations have been exploited in \cite{Zanzi:2015evj} to infer that during a quantum measurement in the MFM, a chameleonic jump from one ground state to another one (where the amount of scale invariance is reduced) must take place. Needless to say, during this jump the harmonic approximation cannot be exploited anymore and strong non linearities must be present in the chameleonic theory. These non linearities break the superposition principle during the (short time of the) measurement.

In this article the diffraction of electrons through a circular hole will be particularly interesting. Let us discuss this experiment from the standpoint of the MFM.

We start considering a plane wave traveling towards a first screen where a small circular hole has been made. The plane wave represents incoming electrons with fixed momentum. It is common knowledge that in this set up, we can use a second screen to perform the measurement of position and the result of the measurement will be a set of circular stripes of interference. Where the interference is constructive, we collect electrons on the second screen. On the contrary, where the interference is destructive, we find no electrons on the second screen. Remarkably, the lagrangian is rotationally invariant but the single electron on the screen breaks the invariance with respect to rotations around the beam axis.

This symmetry breaking is related to the chameleonic nature of the model. When the measurement of position is performed with the second screen, the matter density becomes larger and, as already discussed in \cite{Zanzi:2012ha}, this implies a larger string mass (i.e. a larger UV cut off). In other words, inside the screen, the theory probes short (atomic) distances where rotational symmetry is abundantly broken. The chameleon mechanism links together the energy density of the environment and the ground state. Therefore, the non-symmetric environment at short distances gives us a non-symmetric ground state. This is exactly the condition which defines spontaneous symmetry breaking: we have a lagrangian which is rotationally invariant, but the ground state is not. Now we are ready to discuss the origin of the symmetry breaking related to the single electron on the screen. In \cite{Zanzi:2015evj}, the relativistic formalism of quantum field theory has been linked to the non-relativistic formalism of wave functions: the number density of electrons in the screen can be parametrized exploiting the non relativistic wave function (squared) or the expectation value of the matter field (squared). In this way, the spontaneous symmetry breaking mentioned above is the origin of the symmetry breaking due to the single electron: the selection of one preferred direction by the single electron is analogous to the selection of a preferred angle by a bended rod.

For a detailed discussion of these issues, the reader is referred to \cite{Zanzi:2015evj}.

\section{Quantum mechanics near the big-bang}
\label{QMs}

Near the big-bang the universe is very small and, consequently, quantum effects are particularly relevant. Indeed, the length scales we consider can be comparable to the de Broglie wavelength of particles. Let us apply the ideas of reference \cite{Zanzi:2015evj} near the big-bang.

Let us start considering a radial collapse of the particles towards the black-hole. This radial collapse might be related, for example, to a contracting phase of the universe before the big bang. In particular, let us imagine a beam of particles falling into a spherically symmetric black-hole. The particles are moving radially along the 5th dimension. Classically the particles must fall into the black-hole. However, near the big-bang quantum effects are important: what happens quantum mechanically? When a particle is near the black-hole we can say that the particle enters into the ''shadow'' of the black-hole: the radiation density will be smaller and this shadow can be interpreted as a spherical hole in the energy density of the environment. The beam of particles break the spherical symmetry but an axial symmetry remains around the beam axis. The gravitational field of the black-hole can be interpreted as a screen (orthogonal to the beam direction) where the result of a measurement of position can be collected. Indeed, the effect of a shift in the matter density (a true screen) is the same as a shift in the gravitational field: both effects are summarized by the conformal anomaly and produce a chameleonic jump. This jump from one ground state to a different one produces the collapse of the wave-function in harmony with \cite{Zanzi:2015evj}. The situation is totally analogous to a diffraction experiment of electrons through a circular hole: we know that in a screen orthogonal to the beam axis there will be circular regions (rings) where the particles will be localized and also regions (gaps) where the interference is destructive. The same pattern can be applied in the case of a radial fall into the black-hole. Classically the particles must fall into the black-hole but quantum mechanically particles deviate from the radial direction. Therefore, the particles are localized around the black-hole with the result that only certain circular orbits are allowed by quantum mechanics. Now we can repeat the same argument with a different beam, namely another radial beam but coming from a different angle. We obtained $S^4$ shells of particles where the interference is constructive: these circular orbits correspond to the initial values of the branes' moduli. As already mentioned above, the cosmological evolution of these branes' moduli will be analyzed in a future work. At this stage we simply assume that, after the $S^4$ branes are formed exploiting the quantum interference, the branes' moduli evolve towards the minimum of the Casimir potential.

This scenario has a number of interesting consequences:\\
1) this is a mechanism to localize particles on the $S^4$ branes. As already mentioned in the previous section, when we consider BPS branes in heterotic-M-theory, we can obtain a more realistic model putting matter on branes and detuning the brane tension. This non-BPS configuration is the result, in this article, of a constructive quantum interference.\\
2) The Casimir potential is not only reminiscent of a newtonian effective potential (see \cite{Zanzi:2016het}). The Casimir potential is really a gravitational effective potential for a 2-body problem (even if we are far beyond newtonian physics in this model): one body is the generic particle around the black hole and the second body is the black hole. The $1/r$ term of the black hole is mixed with the $1/r^2$ term related to the angular momentum. We have one Casimir minimum for every brane (see also \cite{Zanzi:2016het}). As already mentioned above, the quantum interference defines the {\it initial} position of the branes and then we assume that these branes evolve dynamically in the minimum of the Casimir potential. A detailed analysis of this cosmological evolution of the branes' moduli will be discussed in a future work.\\
3) This quantum interference pattern is a new way to define branes in heterotic-M-theory. \\

The careful reader might be worried because the analysis of \cite{Zanzi:2015evj} is 4-dimensional but, in this article, we exploit the collapse of the wave function in a 5D set-up. This is not a problem because in the lagrangian of the MFM, the $\Phi$ field of $\phi^2 \Phi^2$ is obtained evaluating the bulk matter field on the brane (see \cite{Zanzi:2016het}). There is no difference between the 4D value of a matter field and the value (on the brane) of the bulk matter field. We infer that when the gravitational field of the black hole becomes strong, there is a jump in the value of the 5D field and this is compatible with the localization of the wave function of the bulk zero mode towards the black hole. The idea that the shadow of the black hole corresponds to a hole in a screen requires only the 4D construction of \cite{Zanzi:2015evj}, because this ''hole'' is orthogonal to the radial direction and, hence, it belongs to the usual 4D dimensions.

From a theoretical point of view, it seems at this stage that the correct quantum framework to discuss the deviation of particles from the radial direction during the collapse is given by Bohmian mechanics. This point must be further elaborated and it is left for a future work.

\section{Modified Schroedinger equation}

The orbifold of time of reference \cite{Zanzi:2016het} is a powerful tool to extract theoretical equations. An interesting example is given by a modified Schroedinger equation for quantum mechanics. There is one major open issue in \cite{Zanzi:2015evj}: the modified Schroedinger equation has not been written. Now we fill the gap.

The orbifold of time is telling us that the field equations in the E-frame are mapped into the field equations in the S-frame. Here is the idea: let us extract the modified Schroedinger equation from the S-frame field equations.
In the S-frame the matter field equation is (see \cite{Fujii:2003pa})
\bea
\Box \Phi -\frac{f}{2} \phi^2 \Phi-\frac{\lambda}{6} \Phi^3 =0.
\eea

We can extract the modified Schroedinger equation following the procedure outlined in \cite{Nikolic:2012wj}\footnote{In particular we can recover the standard Schroedinger equation starting from the Klein-Gordon equation $\Box \Psi +m^2 \Psi=0$
simply writing $\Psi=\frac{e^{-imt}}{\sqrt{m}} \Psi_{NR}$, where $\Psi_{NR}$ is the non-relativistic wave function.}.
We start with a Minkowski approximation which is justified exploiting the Einstein's equivalence principle. Hence we write (the metric signature is $-+++$)

\bea
(-\partial_t^2 + \partial_x^2 -m^2) [\frac{e^{-imt}}{\sqrt{m}} \Psi_{NR}]-\frac{\lambda}{6}  \frac{e^{-3imt}}{\sqrt{m^3}} \Psi^3_{NR}=0
\eea
where the mass is $\phi$-dependent because, by definition, $m^2(\phi)=\frac{f}{2} \phi^2$. Hence, this equation is actually coupled to the field equation for $\phi$.
If we consider the mass roughly constant, we can write the equation as
\bea
\frac{2im}{\sqrt{m}}e^{-imt} \frac{\partial \Psi_{NR}}{\partial t}- \frac{e^{-imt}}{\sqrt{m}} \partial_t^2 \Psi_{NR} + \frac{e^{-imt}}{\sqrt{m}} \nabla^2 \Psi_{NR} - \frac{\lambda e^{-3imt}}{6\sqrt{m^3}}  \Psi_{NR}^3=0.
\eea
To proceed further we make the approximation of reference \cite{Nikolic:2012wj}, namely 
\bea
 \mid \partial_t^2 \Psi_{NR} \mid << m \mid \partial_t \Psi_{NR} \mid.
 \eea
In this way the modified Schroedinger equation is written as:
\bea
i \partial_t \Psi_{NR}= -\frac{\nabla^2}{2m} \Psi_{NR} +\frac{\lambda}{12} \frac{e^{-2imt}}{m^2} \Psi_{NR}^3.
\eea
Remarkably, the equation is highly non-linear in harmony with the chameleonic nature of the theory.  Needless to say, this equation can be linearized using an harmonic approximation near the ground state where the observer lives. The harmonic approximation will break down during the measurement. In our model, this modified Schroedinger equation is valid not only near the big bang but also in our laboratories.

An interesting line of development will search for potential connections with the Yang-Baxter equation.

\section{Conclusions}

We analyzed the role played by quantum mechanics near the big bang. Our scenario is embedded in heterotic-M-theory. The cosmological singularity is avoided because when the universe is small, quantum mechanics (and stringy effects) become important. We focused our attention on a spherical collapse of matter (and radiation) towards a SUGRA-motivated black hole. The collapse of matter (and radiation) takes place radially in the fifth dimension and, in this model, the presence of the spherical shadow of the black hole at short distances is completely analogous to the presence of a spherical hole in a screen of a diffraction experiment with electrons. The final result is that matter does not fall radially into the black hole but quantum effects deviate the particles and create orbits around the black hole which are allowed by quantum mechanics. The chameleonic behaviour of the dilaton before the big bang is guaranteed by the orbifold of time which identifies the pre-big-bang phase with the post-big-bang one. This mechanism to generate $S^4$ branes is interesting for various reasons:\\
1) This is a mechanism to localize particles on the $S^4$ branes and, hence, to leave the BPS configuration. \\
2) The Casimir potential is really a gravitational effective potential for a 2-body problem.\\
3) We obtained a new definition of brane in heterotic-M-theory: the initial position of a brane is the region of space characterized by a constructive quantum interference. The evolution of the branes' moduli towards the Casimir-induced minima will be discussed in a future work.

Another result should be mentioned. Exploiting an orbifold compactification of time, we have written a modified Schroedinger equation (valid also in our laboratories and not only near the big bang). Indeed, the orbifold of time is telling us that the equations valid well before the big bang (like the S-frame field equations) are valid also well after the big bang. This modified Schroedinger equation is strongly non linear but the standard Schroedinger equation is recovered linearizing the theory around the ground state where the observer performs its experiments.

A promising line of development will try to investigate whether it is possible to apply the collapse of the wave function and the quantum interference pattern to study the rings of Saturn and the formation of planetary systems.

%%%%%%%%%%%%%%%%%%%%%%%%%%%%%%%%%%%%%%%%%%%%%%%%%%%%%%%%%%%%%%%%%%%%%%%%%%%%%%%%%%%%%%%%%%%%%%%%%%%%%%%%%%%%%%%%%%%%%
% ACKNOWLEDGEMENTS
%%%%%%%%%%%%%%%%%%%%%%%%%%%%%%%%%%%%%%%%%%%%%%%%%%%%%%%%%%%%%%%%%%%%%%%%%%%%%%%%%%%%%%%%%%%%%%%%%%%%%%%%%%%%%%%%%%%%%
%\vspace{0.5cm}
\subsection*{Acknowledgements}

I am grateful to Massimo Pietroni and Gabriele Veneziano for useful conversations. I warmly thank the Galileo Galilei Institute for Theoretical Physics (Florence) for the kind hospitality: part of this work has been developed at the GGI.
%%%%%%%%%%%%%%%%%%%%%%%%%%%%%%%%%%%%%%%%%%%%%%%%%%%%%%%%%%%%%%

%%%%%%%%%%%%%%%%%%%%%%%%%%%%%%%%%%%%%%%%%%%%%%%%%%%%%%%%%%%%%%%%%%%%%%%%%%%%%%%%%%%%%%%%%%%%%%%%%%%%%%%%%%%%%%%%%%%%%
% BIBLIOGRAPHY
%%%%%%%%%%%%%%%%%%%%%%%%%%%%%%%%%%%%%%%%%%%%%%%%%%%%%%%%%%%%%%%%%%%%%%%%%%%%%%%%%%%%%%%%%%%%%%%%%%%%%%%%%%%%%%%%%%%%%xr}

\providecommand{\href}[2]{#2}\begingroup\raggedright\endgroup

\end{document}